\documentclass[12pt,aaspp4]{aastex}
\usepackage[table]{xcolor}
\usepackage[normalem]{ulem}
\usepackage{rotating}
\begin{document}
\doublespace

\title{A 400 solar mass black hole in the Ultraluminous X-ray source M82 X-1 accreting close to its Eddington limit}
\author{\normalfont{Dheeraj R. Pasham$^1$, Tod E. Strohmayer$^2$, Richard
F. Mushotzky$^1$}} \altaffiltext{1}{Astronomy Department, University of
Maryland, College Park, MD 20742; email: dheeraj@astro.umd.edu;
richard@astro.umd.edu} \altaffiltext{2}{Astrophysics Science Division,
NASA's Goddard Space Flight Center, Greenbelt, MD 20771; email:
tod.strohmayer@nasa.gov}

{\bf 
The brightest X-ray source in M82 has been thought to be an intermediate-mass 
black hole (10$^{2-4}$ solar masses, $M_{\odot}$) because of its extremely high 
luminosity and variability characteristics$^{1-6}$, although some models suggest 
that its mass may be only $\sim$ 20 $M_{\odot}$$^{3,7}$. The previous mass estimates 
are based on scaling relations which use low-frequency characteristic timescales 
which have large intrinsic uncertainties$^{8,9}$. In stellar-mass black holes we 
know that the high frequency quasi-periodic oscillations that occur in a 3:2 ratio 
(100-450 Hz) are stable and scale inversely with black hole mass with a reasonably 
small dispersion$^{10-15}$. The discovery of such stable oscillations thus 
potentially offers an alternative and less ambiguous mass determination for 
intermediate-mass black holes, but has hitherto not been realized. Here, we report 
stable, twin-peak (3:2 frequency ratio) X-ray quasi-periodic oscillations from M82 
X-1 at the frequencies of 3.32$\pm$0.06 Hz and 5.07$\pm$0.06 Hz. Assuming that we 
can scale the stellar-mass relationship, we estimate its black hole mass to be 
428$\pm$105 $M_{\odot}$. In addition, we can estimate the mass using the 
relativistic precession model, from which we get a value of 415$\pm$63 $M_{\odot}$.

}

Oscillations arising from general relativistic effects should scale inversely with 
the black hole mass if they arise from orbital motion near the innermost stable 
circular orbit in the accretion disk$^{16,17}$, and there is observational support 
that they do for stellar-mass black holes$^{10}$ (3-50 $M_{\odot}$). M82 X-1's 
previous mass estimates of a few hundred solar masses combined with the Type-C 
identification$^{2,4,9}$ of its mHz X-ray quasi-periodic oscillations suggest that 
3:2 ratio, twin-peak, high-frequency oscillations analogous to those seen in stellar-mass black holes, 
if present, should be detectable in the frequency range of a few Hz$^{16}$. We 
accordingly searched Rossi X-ray Timing Explorer's ({\it RXTE}'s) proportional counter 
array archival data to look for 3:2 oscillation pairs in the frequency range of 
1-16 Hz which corresponds to a black hole mass range of 50-2000 $M_{\odot}$.

We detected two power spectral peaks at 3.32$\pm$0.06 Hz (coherence, Q = centroid 
frequency ($\nu$)/width($\Delta\nu$) $>$ 27) and 5.07$\pm$0.06 Hz (Q $>$ 40) 
consistent with a 3:2 frequency ratio (Fig. 1a, b). The combined statistical 
significance of the detection is greater than 4.7$\sigma$ (see Methods for 
details). 

The proportional counter array's field of view (1$^{\circ}$$\times$1$^{\circ}$) of 
M82 includes a number of accreting X-ray sources in addition to M82 X-1$^{18}$.
The remarkable stability of the two quasi-periodic oscillations on timescales of a 
few years (Movies 1 \& 2), their 3:2 frequency ratio and their high oscillation 
luminosities strongly suggest they are not low-frequency quasi-periodic oscillations 
from a contaminating stellar-mass black hole (see Methods for details). 
Also a pulsar origin is very unlikely for several reasons. First, a pulsar signal 
would be much more coherent than that of the observed quasi-periodic oscillations, 
which clearly have a finite width. Second, based on the observed high quasi-periodic 
oscillation luminosities it is extremely implausible that they originate from a 
pulsar (see Methods for details). Finally, it would be highly coincidental 
to have two pulsars in the same field of view with spins in the 3:2 ratio. Also, based 
on the average power spectra of the background sky and a sample of accreting super-massive  
black holes monitored by the proportional counter array in the same epoch as M82, we
rule out an instrumental origin for these oscillations (Extended Data Figs. 2, 3 
\& 4). This leaves M82 X-1, persistently the brightest 
source in the field of view, as the most likely source associated with the 3:2 ratio 
quasi-periodic oscillation pair.

We estimated M82 X-1's black hole mass, assuming the 1/$M$ (mass) scaling of stellar-mass 
black holes and the relativistic precession model$^{19,20}$, to be 428$\pm$105 
$M_{\odot}$ and 415$\pm$63 $M_{\odot}$, respectively (Fig. 2a,b). Combining the 
average 2-10 keV X-ray luminosity$^{21,22}$ of the source of 5$\times$10$^{40}$ 
ergs s$^{-1}$ with the measured mass suggests that the source is accreting 
close to the Eddington limit with an accretion efficiency of 0.8$\pm$0.2.

The previous mass measurements of M82 X-1 have large uncertainties owing to both 
systematic and measurement errors. For example, modeling of its X-ray energy spectra 
during the thermal-dominant state using a fully relativistic multi-colored disk model 
suggests that it hosts an intermediate-mass black hole with mass anywhere in the range 
of 200-800 $M_{\odot}$ and accreting near the Eddington limit$^{3}$. In addition to the 
large mass uncertainty associated with the modeling, the same study also found that the 
energy spectra can be equally well-fit with a stellar-mass black hole accreting at a rate   
of roughly 160 higher than the Eddington limit$^{3}$. Also, the X-rays from this source are 
known to modulate with a periodicity of 62 days which has been argued to be the orbital 
period of an intermediate-mass black hole$^{22}$--formed in a nearby star cluster MCG 11 
by stellar runaway collisions$^{5,23,24}$--accreting matter via Roche lobe overflow from a 22-25 
$M_{\odot}$ companion star$^{24}$. Detailed stellar binary evolution simulations suggest that the long 
periodicity is best explained by an intermediate-mass black hole with mass in the range of 
200-5000 $M_{\odot}$$^{5,24}$. But, a recent study finds that this periodicity may instead be due to 
a precessing accretion disk in which case a stellar-mass black hole will suffice to explain 
the apparent long periodicity$^{21}$.  

One of the main lines of argument for an intermediate-mass black hole in M82 X-1 is by inverse 
mass scaling of its mHz quasi-periodic oscillations (frequency range of 37-210 mHz$^{9,25}$; 
Fig. 1c) to the Type-C low-frequency X-ray oscillations of stellar-mass black holes 
(frequency range of 0.2-15 Hz)$^{10,26}$. There are two uncertainties with such scaling: (1) it was 
unclear--until now--whether these mHz oscillations are indeed the Type-C analogs of stellar-mass 
black holes$^{8,9}$ and (2) both the Type-C and the mHz oscillations are variable, resulting in a large dispersion 
in the measured mass of 25-1300$M_{\odot}$$^{1,2,4,6}$. The discovery of a stable 3:2 high-frequency periodicity simultaneously 
with the low-frequency mHz oscillations allows for the first time to set the overall frequency scale 
of the X-ray power spectrum. This result not only asserts that the mHz quasi-periodic oscillations of M82 
X-1 are the Type-C analogs of stellar-mass black holes but also provides an independent and the most  
accurate black hole mass measurement to-date. 

Finally, it should be pointed out that while the rms amplitudes (3-5\%) of the oscillations reported here  
and their frequencies with respect to the mHz oscillations (a factor of a few 10s higher) are similar to those 
observed in stellar-mass black holes$^{10}$, they appear narrower (with Q values of 27, 40 \& 80) compared 
to stellar-mass black holes which have Q values of $<$ 20$^{14,30}$. In stellar-mass black holes the Q factor appears 
to be energy dependent in some cases$^{14}$ and it is plausible that a similar effect may be operating 
in this case.

\begin{figure*}
  \begin{center}
 \includegraphics[width=18cm]{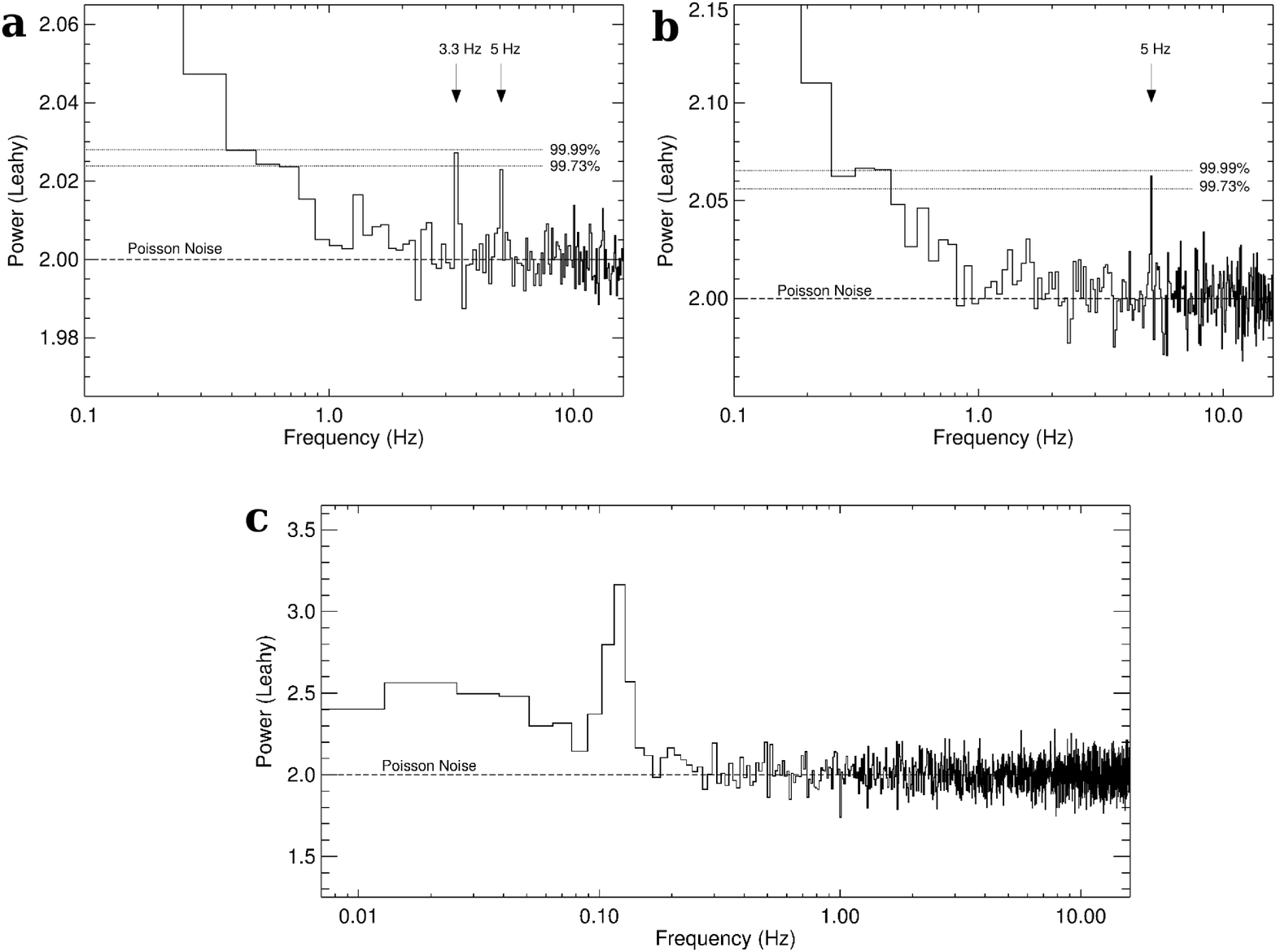} 
  \end{center}
         {\textbf{Figure 1 $\vert$ Power density spectra of M82.} {\bf (a)} Six-year average X-ray (3-13
keV) power density spectrum of M82 using 128-second ($\times$7362) individual light curves. The 
frequency resolution is 0.125 Hz. The two strongest features in the power spectrum occur at 
3.32$\pm$0.06 and 5.07$\pm$0.06 Hz consistent with a 3:2 frequency ratio. {\bf (b)} Averaged 
power density spectrum of all 1024-second ($\times$363) segments. The frequency resolution is 
0.0625 Hz. The strongest feature is at 5 Hz. {\bf (c)} For a direct comparison with stellar-mass 
black holes, we show the broadband power density spectrum of M82 (using $\approx$ 100 ksecs of {\it XMM-Newton}/EPIC 
data; ID: 0206080101) showing the low-frequency quasi-periodic oscillation at 120 mHz in addition to the 
high-frequency quasi-periodic oscillation pair in the top panels.}
     \label{fig:fig2}
\end{figure*}

\begin{figure*}
  \begin{center}
 \includegraphics[width=8cm]{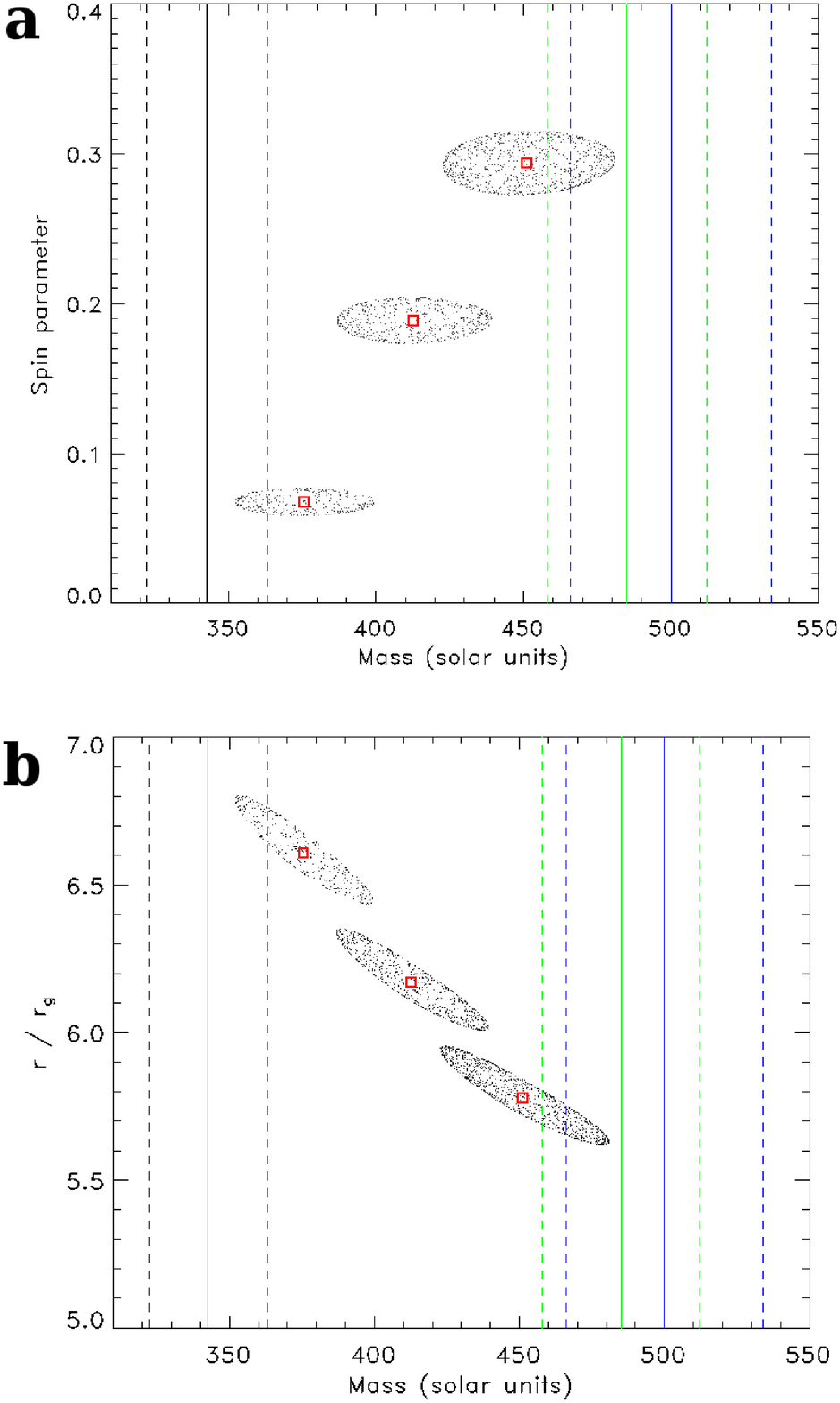}  
 \end{center}
        {\textbf{Figure 2 $\vert$ Mass, spin and radius measurements.}  {\bf (a)} Contours (90\% confidence) 
of M82 X-1's mass as a function of the spin parameter. The three contours correspond to the three 
low-frequency values (37 mHz, 120 mHz and 210 mHz) with the mass increasing as the low-frequency oscillation 
frequency increases. {\bf (b)} Contours of M82 X-1's mass as a function of the radius of the origin 
of these oscillations (in units of $r_{g}$ = $GM/c^{2}$, where $G$, $M$, $c$ are the Gravitational 
constant, the black hole mass and the speed of light, respectively). In both the panels the vertical lines (solid: 
solution; dashed: upper-lower limits) represent M82 X-1's mass estimates assuming a simple
1/$M$ scaling for the high-frequency quasi-periodic oscillations. The three colors correspond to 
scalings using the three microquasars (green: GRO J1655-40$^{27}$; blue: XTE J1550-64$^{28}$; black: 
GRS 1915+105$^{29}$).}
     \label{fig:fig2}
\end{figure*}

\begin{figure*}
  \begin{center}
 \includegraphics[width=16cm]{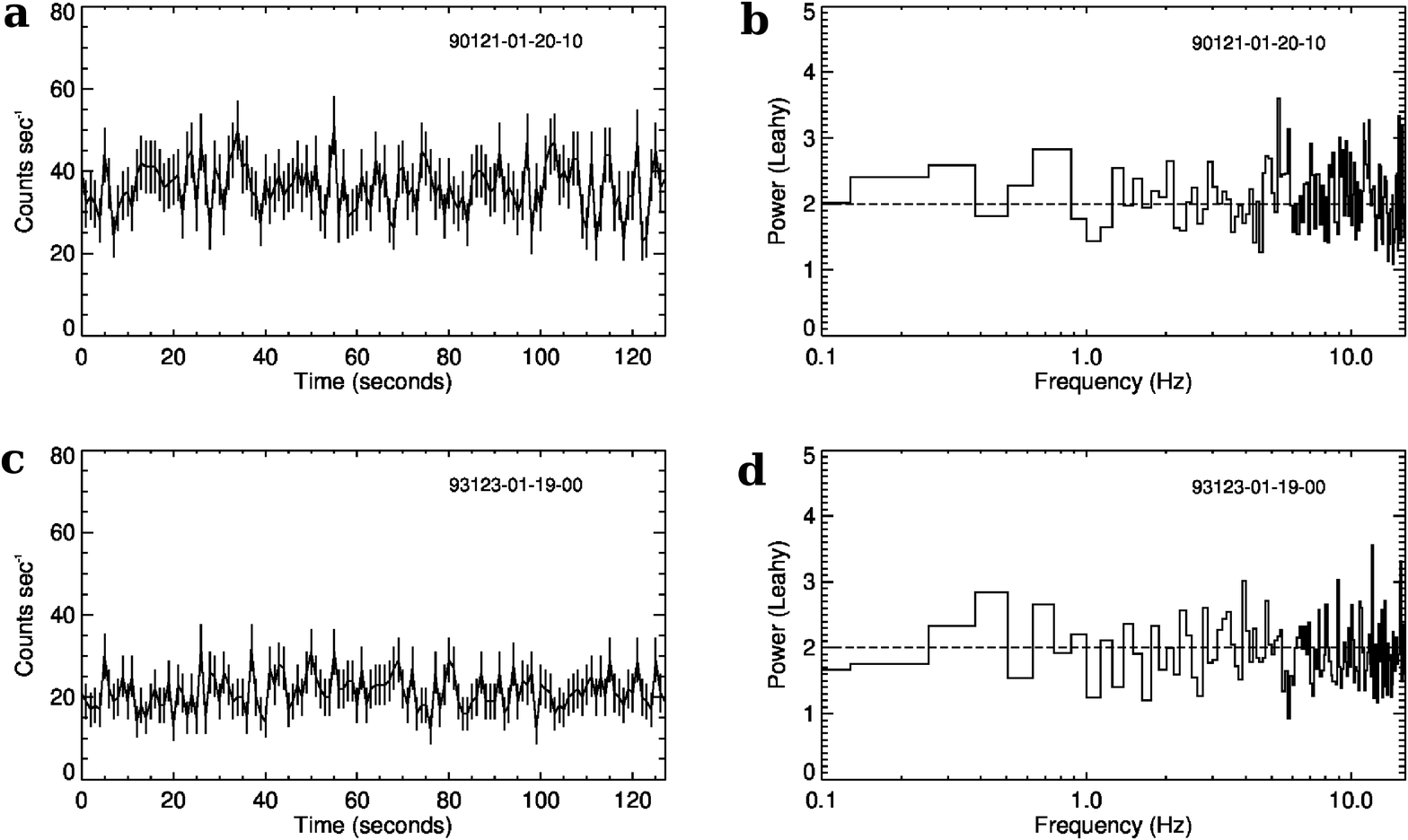}   
  \end{center} 
       {\textbf{Extended Data Figure 1 $\vert$ Sample {\it RXTE} 
proportional counter array light curves and power density spectra of M82. } The 128-second X-ray (3-13 keV)
light curves (a, c) and their corresponding power spectra (b, d) of M82. The corresponding observation IDs are shown in
the top right of each panel. The light curves have a bin size of 1
second while the power spectra have a frequency resolution of 0.125
Hz. The errorbars in (a, c) represent the standard error on the mean.}
     \label{fig:fig2}
\end{figure*}

\begin{figure*}
  \begin{center}
 \includegraphics[width=13cm]{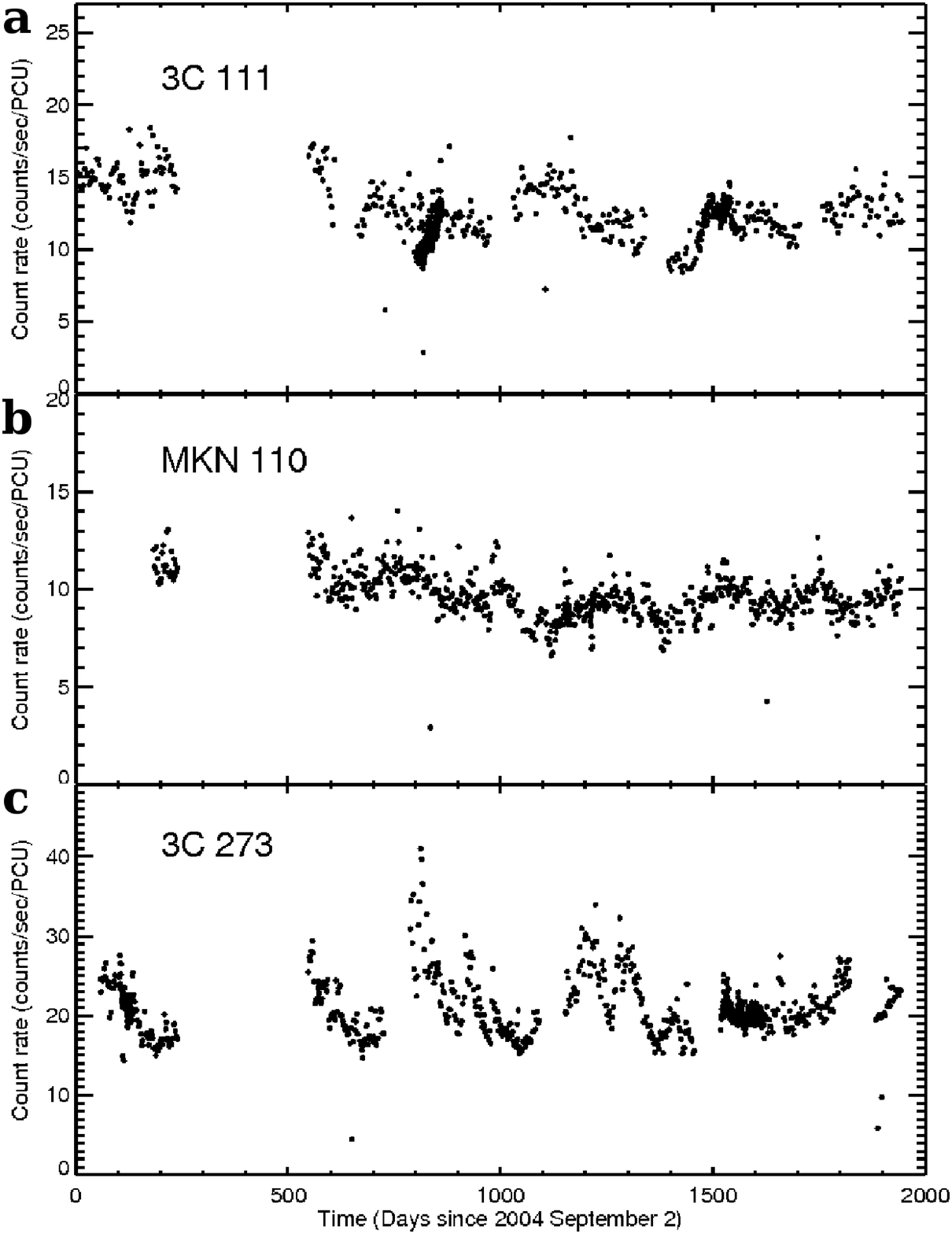}    
        \end{center}
{\textbf{Extended Data Figure 2 $\vert$ Long-term X-ray (3-13 keV) light curves of three accreting super-massive black holes.} These were extracted from the same time window as M82 observations (2004 September 02 -- 2005 April 30 and 2006 March 03 -- 2009 December 30). The name of the galaxy is indicated in the top left of each panel. The count rates are not corrected for background. }
     \label{fig:fig2}
\end{figure*}

\begin{figure*}
  \begin{center}
 \includegraphics[width=10cm]{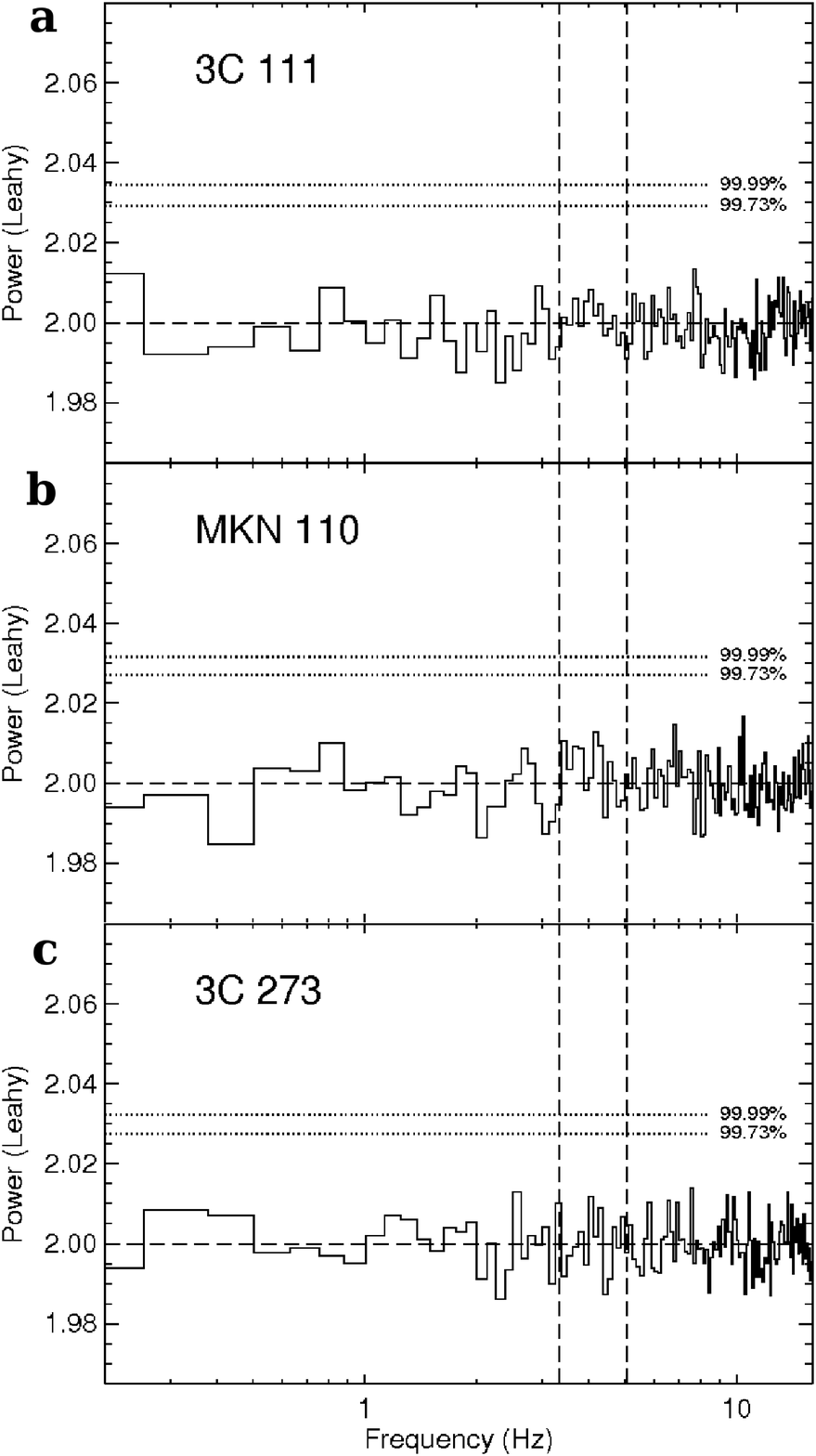}    
        \end{center}
{ \textbf{Extended Data Figure 3 $\vert$ Average X-ray (3-13 keV) power spectra of three accreting super-massive black holes.} Similar to the M82 analysis, these spectra were extracted by combining all the data (128-second data segments) shown in Extended Data Fig. 2. The Poisson noise level is equal to 2 while the 99.73\% and 99.99\% confidence contours are indicated by horizontal dotted lines. The two dashed vertical lines are drawn at 3.32 Hz and 5.07 Hz. Clearly, there are no significant power spectral features. The total PCA exposures used for these spectra were 630 ksecs (top), 738 ksecs (middle) and 711 ksecs. }
     \label{fig:fig2}
\end{figure*}

\begin{figure*}
  \begin{center}
 \includegraphics[width=12cm]{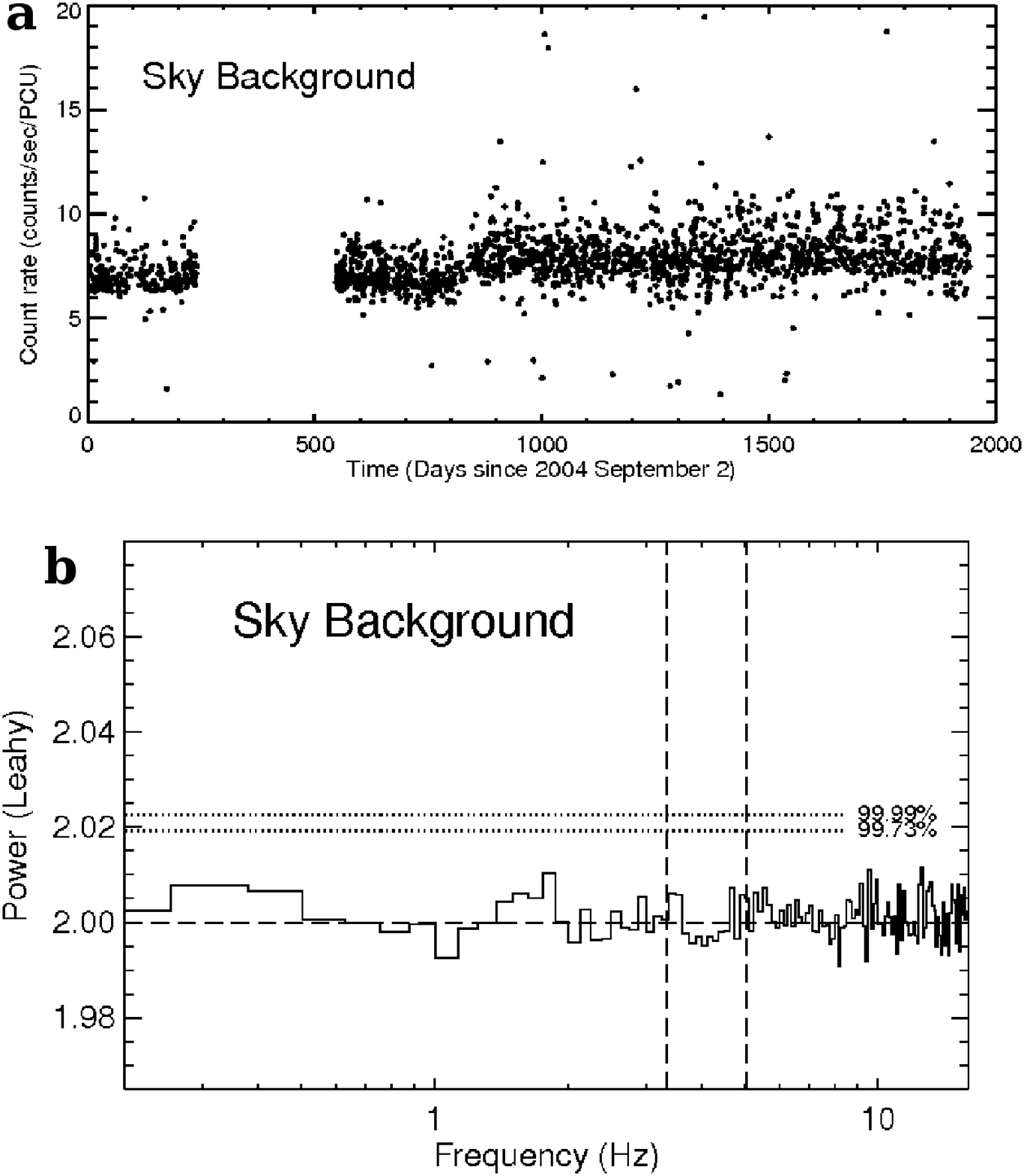}
 \end{center}    
     { \textbf{Extended Data Figure 4 $\vert$ Long-term X-ray (3-13 keV) light curve (top) and the average power density spectrum (bottom) of a background sky field (ra=5.0 deg, dec = -67.0 deg) as observed by {\textit RXTE}/PCA.} Similar to the Extended Data Fig. 3 the 99.73\% and 99.99\% contours and the vertical lines at 3.32 and 5.07 Hz are indicated. Again only observations coincident with M82 monitoring were used (2004 September 02 -- 2005 April 30 and 2006 March 03 -- 2009 December 30) and the total exposure time was 1450 ksecs. }
     \label{fig:fig2}
\end{figure*}

\newpage


\noindent {\bf Acknowledgements } This work is based on observations made with 
the Rossi X-ray Timing Explorer ({\it RXTE}), a mission that was managed and controlled 
by NASA's Goddard Space Flight Center (GSFC) in Greenbelt, Maryland, 
USA. All the data used in the present article is publicly available 
through NASA's {\it HEASARC} archive. Pasham would like to thank Margaret Trippe, Cole 
Miller and Chris Reynolds for valuable discussions. Finally, we would like to thank 
the referee for suggesting to check for pulsations in PCA's AGN data.

\noindent {\bf Author contributions } Dheeraj R. Pasham and Tod E. Strohmayer both reduced the data and carried out the analysis. They also wrote the paper. Richard F. Mushotzky contributed towards the interpretation of the results.

\noindent {\bf Author information } Reprints and permissions information is available at www.nature.com$/$reprints. The authors declare no competing financial interests. Correspondence and requests for materials should be addressed to dheeraj@astro.umd.edu.

\newpage

\noindent{\Large {\bf Methods}}

\noindent{\bf Estimating the expected significance of the quasi-periodic oscillations.}
The detectability (statistical significance, $n_{\sigma}$)
of a quasi-periodic oscillation feature can be expressed as,
\begin{equation}
n_{\sigma} = \frac{1}{2} r^2 \frac{S^2}{(S + B)} \sqrt{\frac{T} {\Delta\nu}} \; ,
\end{equation}
where $r$, $S$, $B$, $T$, and $\Delta\nu$ are the rms
(root-mean-squared) amplitude of the quasi-periodic oscillation, the source count rate, the
background count rate, the exposure time, and the width of the quasi-periodic oscillation,
respectively$^{31}$. Assuming an rms amplitude of a few
percent--similar to that seen in stellar-mass black holes$^{30}$--and using the mean {\it
RXTE} proportional counter array source and background rates obtained from prior observations$^{21}$, 
we found that the wealth of publicly
available, archival {\it RXTE} monitoring data ($\approx$ 1
Megasecond) spread across a timespan of $\approx 6$ years would
provide a sensitive search for high-frequency quasi-periodic oscillations in M82 X-1.

\noindent{\bf Data primer.}
M82 was monitored (0.5-2 ksecs roughly once every three days) with the Rossi X-ray 
Timing Explorer's ({\it RXTE}'s) proportional counter array ({\it PCA}) between 
1997 February 2 until 1997 November 25 (0.8 years) and 2004 September 2 until 2009
December 30 (5.3 years).  All the {\it PCA} observations were carried
out in the {\it GoodXenon} data acquisition mode. The total number
of monitoring observations was 867, which were divided amongst six
proposals ({\it RXTE} proposal IDs: P20303, P90121, P90171, P92098,
P93123, P94123). As recommended by the data analysis guide provided 
by the {\it RXTE} Guest Observer Facility (GOF)
(https://heasarc.gsfc.nasa.gov/docs/xte/abc/screening.html), we first
screened the data to only include time intervals that satisfy the
following criteria: ELV $>$ 10.0 \&\& OFFSET $<$ 0.02 \&\&
(TIME\_SINCE\_SAA $<$ 0 $||$ TIME\_SINCE\_SAA $>$ 30) \&\& ELECTRON2
$<$ 0.1. In addition to the above standard filters, we only used X-ray
events within the energy range of 3-13 keV which translates to {\it
PCA} channels 7-32. This energy range is comparable to the bandpass in
which high-frequency quasi-periodic oscillations have been reported from stellar-mass black holes$^{10,14,15}$. Moreover, beyond 13 keV
the background dominates the overall count rate by a factor greater
than 10.  For each observation we used all active proportional counter
units (PCUs) in order to maximize the count rate, and thus the 
sensitivity to quasi-periodic oscillations.

Before we extracted the individual power spectra we created individual light curves
(using a bin size of 1 second) from all the observations. Through
manual inspection we removed a small number of observations affected
by flares, as these are attributed to background events not associated
with the source, for example, gamma-ray bursts. Extended Data Fig. 1 shows two sample 
light curves and their corresponding power spectra. They represent 
the typical quality of the individual light curves used in extracting the average power spectra.

\noindent {\bf Estimating the statistical significances.}
We first divided the data into 128-second segments and extracted 
their light curves with a time resolution of 1/32 seconds. We then constructed a 
Leahy normalized$^{32}$ power density spectrum (where the Poisson noise level 
is 2) from each of these 128-second light curves. All the power spectra were then 
combined (7362 individual power spectra) to obtain a six-year averaged power density spectrum of M82 (Fig. 1a). 

In order to estimate the statistical significance of any features in
the 1-16 Hz range of the six-year averaged power density spectrum obtained using the 128-second 
data segments, we first ensured that
the local mean was equal to 2, the value expected from a purely
Poisson (white noise) process. We then computed the probability, at
the 99.73\% (3$\sigma$) and the 99.99\% (3.9$\sigma$) confidence
levels, of obtaining the power, P = P$_{*}$$\times$7362$\times$16 from
a $\chi^2$ distribution with 2$\times$7362$\times$16 degrees of
freedom. Here P$_{*}$ is the power value of a statistical fluctuation
at a given confidence level. We used this $\chi^2$ distribution
because we averaged in frequency by a factor of 16 and averaged 7362
individual power spectra. Considering the total number of trials
(frequency bins within 1-16 Hz) we computed the 99.73\%
(1/(371$\times$trials)) and the 99.99\% (1/(10000$\times$trials))
confidence limits (horizontal dotted lines in Fig. 1a). We detect two power spectral 
peaks at 3.32$\pm$0.06 Hz and 5.07$\pm$0.06 Hz significant at the 2$\times$10$^{-4}$ (3.7$\sigma$) 
and 6$\times$10$^{-3}$ (2.75$\sigma$) levels, respectively, assuming both features were 
searched for independently between 1-16 Hz. However, after identifying the first feature at 3.3 Hz, 
if we are searching for a second feature at a 3:2 ratio, then the search from thereon 
only includes the bins nearby 3/2 or 2/3 of 3.3 Hz. If this is taken into consideration 
the significance of the 5 Hz feature increases due to the smaller number of trials to 
5$\times$10$^{-4}$ or 3.5$\sigma$. In order to further test the significance of the 
5 Hz feature, we extracted an average power density spectrum using all the data with segments longer than
1024 seconds (Fig. 1b). The 
5 Hz feature (Q $>$ 80) is clearly detected at the 1.5$\times$10$^{-4}$ confidence level or
3.8$\sigma$, considering a full search between 1-16 Hz.

The combined probability of two independent chance fluctuations--in the 3:2 frequency 
ratio--one at the 3.7$\sigma$ level (3.3 Hz feature) and the other at the 2.75$\sigma$ 
level (5 Hz feature) is greater than 4.7$\sigma$. 

\noindent{\bf Signal cannot be from a single observation.}
The very presence of these two features in the six-year averaged power 
spectrum suggests that they are stable on this timescale.
In order to further rule out the possibility that these oscillations are due to a
single or small number of particular observations, we constructed two
dynamic average power spectra, one for the 128-second segments
(dynamic power density spectrum\#1 or Movie 1) and another for the 1024-second segments (dynamic
power density spectrum\#2 or Movie 2).  These track the evolution of the average power density spectra as a function
of the total number of individual power spectra used in constructing
the average. The two dynamic power density spectra clearly suggest that the power in
these two features builds-up gradually as more data is being averaged,
as opposed to a sudden appearance, which would be expected if a single
or a small number of observations were contributing all the signal
power. In addition, dynamic power density spectrum\#1 clearly shows that while the 5 Hz
feature is stronger during the earlier stages of the monitoring
program the 3.3 Hz feature is stronger during the later
observations. Longer exposures of the order of 1-2 ksecs were carried
out during the earlier stage of the monitoring program, which explains
the higher significance of the 5 Hz feature in the average power density spectrum of the
1024-second segments (Fig. 1b).

\noindent{\bf RMS amplitude of the quasi-periodic oscillations.}
To calculate the rms variability amplitude of these quasi-periodic oscillations we first
determined the mean net count rate (source + background) of all the
light curves used to extract the average power spectra. These values
were equal to 29.9 counts s$^{-1}$ and 35.2 counts s$^{-1}$ for the
128-second and the 1024-second segment power spectra,
respectively. The rms amplitudes of the 3.3 Hz and the 5 Hz quasi-periodic oscillations, not
correcting for the background, were estimated to be 1.1$\pm$0.1\% and
1.0$\pm$0.1\%, respectively. Similarly, the rms amplitude of the 5 Hz
feature within the 1024-second segment power density spectrum was estimated to be
1.1$\pm$0.1\%. We then estimated the mean background count rate from
the {\it Standard2} data utilizing the latest {\it PCA} background
model. The mean background rates during the 128-second and the
1024-second segments were estimated to be 18.9 counts s$^{-1}$ and
24.0 counts s$^{-1}$, respectively. After accounting for the X-ray
background we find that the rms amplitudes of the 3.3 Hz and the 5 Hz
features--averaged over the entire data--are 3.0$\pm$0.4\% and
2.7$\pm$0.4\%, respectively, while the amplitude of the 5 Hz feature
within the 1024-second power density spectrum was estimated to be 3.5$\pm$0.4\%.

Furthermore, the source count rates estimated above ({\it net} minus
{\it background}) represent the combined contribution from all the
X-ray point sources within the {\it PCA}'s $1^{\circ}$$\times$$1^{\circ}$ field of
view$^{18}$. Thus, the quasi-periodic oscillation rms amplitudes are underestimated. A study using
the high-resolution camera (HRC) on board {\it Chandra} suggests that
there are multiple point sources within the 1'$\times$1' region around
M82 X-1$^{18,33}$. Tracking the
long-term variability of these sources suggests that the maximum
luminosity reached by any of these sources--except for source 5 (as
defined by ref. 18)--is less than 1/5$^{th}$ of the
average luminosity of M82 X-1$^{33}$. Source 5 is a
highly variable transient ULX with its 0.5-10 keV luminosity varying
between 10$^{37-40}$ ergs s$^{-1}$ (see the middle-left panel of 
Fig. 1 of ref. 33). The quasi-periodic oscillations reported here are most likely produced from M82 X-1,
which has persistently been the brightest source of any in the
immediate vicinity of M82 X-1 (see the following sections). Although a precise value
of the rms amplitude cannot be evaluated using the current data, we
estimate an absolute upper limit by calculating the inverse of the
fraction of the count rate contribution from M82 X-1, assuming all the
remaining contaminating sources are at their brightest ever
detected. This scenario is highly unlikely but will serve as an
absolute upper bound to the rms amplitudes of the quasi-periodic oscillations, assuming they
are from M82 X-1.  Using the values reported by ref. 33 
the fraction is roughly 1.8. Thus the
true rms amplitudes of the 3.3 Hz and the 5 Hz quasi-periodic oscillations are estimated to
be in the range of 3-5\%.

Also, {\it XMM-Newton}'s EPIC instruments -- with an effective area
of $\approx$ 1/5$^{th}$ of {\it RXTE}'s {\it PCA} albeit with lower
background -- observed M82 on multiple epochs, with a total effective
exposure of $\approx$ 350 ksecs. These observations were taken in the
so-called full-frame data acquisition mode with a time resolution of
73.4 ms or a Nyquist frequency of 6.82 Hz. This value is close to the
quasi-periodic oscillation frequencies of interest and causes some signal
suppression$^{31}$. Nevertheless, we extracted an average 3-10 keV power density spectrum with
128-second data segments using all the observations (2718 individual
power spectra). We do not detect any statistically significant
features nearby 3.3 and 5 Hz, however, we estimate a quasi-periodic oscillation upper limit
(3$\sigma$ confidence) of 5.2 and 6.2\% rms (using Eq. 4.4 and
Eq. 4.10 of ref. 28) at 3.3 and 5 Hz, respectively, which
are roughly twice the rms values of the quasi-periodic oscillations detected in the {\it PCA}
data.

Using the 128-second data segments from {\it RXTE}, we also studied the energy dependence 
of the rms amplitudes of the two oscillations (see Extended Data Table 1). While the 
error bars are large, there appears to be a modest decrease in the rms amplitudes of 
these oscillations at lower X-ray energies. {\it XMM-Newton}'s EPIC instruments are 
more sensitive in the 3-8 keV band which is comparable to the PCA channels of 7-18 
(the first and the fourth rows of the Extended Data Table 1). 

\noindent {\bf Ruling out low-frequency quasi-periodic oscillations.}
Low-frequency quasi-periodic oscillations of stellar-mass black holes, such as the
Type-C quasi-periodic oscillations$^{26}$, have typical centroid frequencies of a
few Hz with rms amplitudes$^{26}$ of 5-25\%, but are known to vary in
frequency by factors of 8-10 over time scales of days$^{34,35}$. 
 This would lead to very broad features in the kind of
average power spectra we have computed from. Moreover, among the plethora of low-frequency quasi-periodic oscillations
currently known there is no indication of them preferentially
occurring with a 3:2 frequency ratio. Furthermore, the average
luminosity of the quasi-periodic oscillations reported here is $\approx$ 0.03$\times$(the
average luminosity of all the sources observed by the {\it PCA} in the
3-13 keV band)$^{21,22}$ which is $\approx$ 0.03$\times$5$\times$10$^{40}$ ergs
s$^{-1}$ = 1.5$\times$10$^{39}$ ergs s$^{-1}$. This is comparable to or more than the peak
X-ray luminosities of the contaminating sources, except for source 5$^{33,36}$. 
Therefore, if these features were simply low-frequency quasi-periodic oscillations
produced by any of the contaminating sources--except for source
5--their X-ray flux would have to be modulated at almost 100\%, which
is not plausible for the typical amplitudes of low-frequency quasi-periodic oscillations.

Source 5, which is a ULX, could in principle be the
origin of these 3:2 ratio quasi-periodic oscillations.  However, 3-4 mHz quasi-periodic oscillations have been
discovered from this source and have been identified as Type-A/B quasi-periodic oscillations
analogs of stellar-mass black holes$^{37}$.  Such a characterization for the
mHz quasi-periodic oscillations suggests that the ULX might host a black hole with a mass of
12,000-43,000 $M_{\odot}$ $^{37}$. If that were the case,
the expected frequency range of high-frequency quasi-periodic oscillation analogs for a few 10,000
$M_{\odot}$ black hole would be a few 100s of mHz, a factor of 10 lower than 
the quasi-periodic oscillations reported here, thus suggesting that
the 3.3 and 5 Hz quasi-periodic oscillations are less likely to be the high-frequency quasi-periodic oscillation analogs of
source 5. 

\noindent{\bf Ruling out pulsar origin.}
Rotation-powered pulsars can be strongly excluded, they simply cannot 
provide the required luminosity. A neutron star's rotational energy 
loss rate can be expressed in terms of its moment of inertia, $I $, 
spin period $P$, and period derivative, $\dot P$ as,
\begin{equation}
\dot E_{rot} = \frac{2\pi^2 I \dot P}{P^3}.
\end{equation}
No known pulsar has a spin-down luminosity comparable to the estimated
quasi-periodic oscillation X-ray luminosity.  For example, the energetic Crab pulsar has
$\dot E \approx 2\times 10^{38}$ ergs s$^{-1}$, and only a fraction of
a pulsar's spin-down power typically appears as X-ray radiation. This
rules out rotation-powered pulsars. As M82 is a starburst galaxy it
likely hosts a population of accreting X-ray pulsars. Such
accretion-powered pulsar systems are typically limited by the
Eddington limit of $\approx 2\times10^{38}$ ergs s$^{-1}$ for a
``canonical" neutron star. Useful comparisons can be made with the
population observed with the {\it RXTE/PCA} in the Small Magellanic
Cloud (SMC)$^{38}$. These authors present pulsed
luminosities for the SMC pulsar population, and none is larger than
$\approx 3 \times 10^{38}$ ergs s$^{-1}$. Again, this is much smaller
than the inferred quasi-periodic oscillations luminosities.  Moreover, such pulsars are
variable, and their time-averaged luminosities would be reduced
further by their outburst duty cycles. At present the only pulsar that
is known to reach luminosities of $\sim$ 10$^{40}$ ergs s$^{-1}$ for
brief periods of time is GRO J1744-28--the so-called bursting pulsar$^{39-41}$. This
object has a 2.1 Hz spin frequency and was discovered during an
outburst that spanned the first 3 months of 1996 (we note that at the
time of writing the source was detected in outburst again, suggesting
a duty cycle of about 18 years, ATel \#5790, \#5810, \#5845, \#5858,
\#5883, \#5901).  It's peak persistent luminosity (assuming a distance
close to that of the Galactic center) was $\approx 7 \times 10^{38}$
ergs s$^{-1}$.  With a pulsed amplitude of about 10\% this would still
give a pulsed luminosity much less than the inferred luminosities of the quasi-periodic oscillations.
The Type II--accretion driven--bursts from this source$^{42}$ could reach
about $10^{40}$ ergs s$^{-1}$, and with a 10\%
pulsed amplitude this could give an instantaneous luminosity close to
that of the average quasi-periodic oscillation luminosities. However, the bursting intervals
make up less than 1\% of the total time, and thus this small duty
cycle will reduce the average pulsed luminosity due to the bursts to a
level substantially below that of the observed quasi-periodic oscillations. Thus, we conclude that
the observed quasi-periodic oscillations cannot be associated with accreting pulsars in M82.
 
\noindent{\bf Ruling out instrumental origin.}
We also rule out the possibility that this signal is intrinsic to {\it RXTE}/PCA by 
extracting the average power density spectra of a sample of accreting super-massive black 
holes with PCA count rates comparable to M82. To be consistent we only used monitoring 
data taken in the {\it GoodXenon} mode during the same epoch as M82. The long-term 
light curves of these sources in the same time range as M82 are shown in the Extended Data 
Fig. 2. Based on the causality argument active galactic nuclei with black hole masses 
greater than 10$^{6}$ $M_{\odot}$ cannot have coherent oscillations at frequencies 
higher than $\approx$ 0.1 Hz. The average power spectra obtained with PCA in the 3-13 keV
bandpass are essentially flat and are consistent with being Poisson noise (see Extended Data Fig. 
3). In addition, we also extracted the average power spectrum of a blank sky field (background) 
monitored using the {\it GoodXenon} mode during the same epoch as M82. The corresponding 
average power spectrum is shown in the Extended Data Fig. 4b and is again consistent with 
being featureless, as expected. 

\noindent {\bf Relativistic precession model analysis.}
Under the interpretation of the relativistic precession model (RPM), 
the upper harmonic of the high-frequency quasi-periodic oscillation is associated with the Keplerian 
frequency at some inner radius while the lower harmonic of the high-frequency quasi-periodic oscillation 
and the Type-C quasi-periodic oscillation are associated with the periastron and nodal 
precession frequencies, respectively, at the same radius. Recently, 
ref. 20 have applied this model to GRO J1655-40 which 
exhibits both the low-frequency and the high-frequency quasi-periodic oscillations and has 
a very accurate mass measurement of $5.4\pm 0.3 M_{\odot}$$^{27}$. 
They find that the black hole mass evaluated from the relativistic precession model analysis 
agrees nicely with its dynamical mass estimate. Given this promise 
of the relativistic precession model, we estimated the mass and spin of M82 X-1's black hole 
using this model and essentially following the methodology as in ref. 20. 

The relativistic precession model analysis requires that the three quasi-periodic oscillations, the
two high-frequency quasi-periodic oscillations and a low-frequency quasi-periodic oscillation, be 
observed simultaneously. This is however not the case for M82 observations. While the combined six-year
{\it RXTE's} proportional counter array data shows the twin high-frequency quasi-periodic oscillation 
pair, individual {\it XMM-Newton} observations randomly dispersed over the same epoch as the {\it RXTE} 
monitoring have shown mHz low-frequency quasi-periodic oscillations with frequencies in the range of 
37-210 mHz (see Table 2 of ref. 10 and ref. 23). Thus, we carried out the analysis for three separate 
values of the low-frequency quasi-periodic oscillations, the lowest and the highest values of 37 mHz 
and 210 mHz, respectively, as well as a mean quasi-periodic oscillation frequency of 120 mHz.

The dimensionless spin parameter is constrained 
to the range $0.06 < a < 0.31$, and the inferred radius, $r$, in 
the disk is in the range $5.53 < r/r_g < 6.82$, where $r_g = GM/c^2$ (G and c are the gravitational constant
and the speed of light, respectively, while $M$ is the mass of the black hole).


\newpage
\begin{flushleft}
{\scriptsize
 \textbf{Extended Data Table 1 $\vert$ Dependence of the \% rms amplitudes of the two oscillations on X-ray bandpass$^{\dagger}$. }
\begin{tabular}{ m{1.5cm} m{1.5cm} m{1.5cm} m{1.5cm} m{2.1cm} m{2.1cm} }
 \multicolumn{6}{c}{} \\ [4pt]
 \hline \\ [-12pt]
PCA Channels 		& Energy Range	& Net Count Rate$^{a}$		& Background Count Rate		 & Uncorrected	\% rms Amplitude$^{b}$	& Corrected \% rms Amplitude$^{c}$ \\ [4pt]
 \hline \\ [-12pt]
 \multicolumn{6}{c}{3.32 Hz Quasi-Periodic Oscillation } \\ [4pt]
 \hline \\ [-12pt]
7-18 			& 3-8 keV  	&  19.0  			& 10.5   		 & 1.1$\pm$0.2  	&  2.8$\pm$0.4    \\
\\[-12pt]
7-24			& 3-10 keV  	&  24.3   			& 14.5   		 & 1.1$\pm$0.2  	&  2.5$\pm$0.4      \\ 
\\[-12pt]
7-32			& 3-13 keV   	&  29.9  			& 18.9   		 & 1.1$\pm$0.1		&  2.7$\pm$0.4       \\ 
\\[-12pt]

\hline \\ [-12pt]
 \multicolumn{6}{c}{5.07 Hz Quasi-Periodic Oscillation } \\ [4pt]
 \hline \\ [-12pt]
7-18 			& 3-8 keV  	&  19.0  			& 10.5   		 & 1.2$\pm$0.2  	&  2.5$\pm$0.4    \\
\\[-12pt]
7-24			& 3-10 keV  	&  24.3   			& 14.5   		 & 1.0$\pm$0.2  	&  2.8$\pm$0.4      \\ 
\\[-12pt]
7-32			& 3-13 keV   	&  29.9  			& 18.9   		 & 1.0$\pm$0.1		&  3.0$\pm$0.4       \\ 
\\[-12pt]
 
 \hline \\
\end{tabular}
\newline
$^{\dagger}${We used 128-second data segments for this study.}
$^{a}${The total (source + background) count rate in the given energy range.}
$^{b}${Not corrected for the background.}
$^{c}${Background corrected \% rms amplitude where corrected rms amplitude = (uncorrected rms amplitude)$\times$(Total Count Rate)/(Source Count Rate). The source count rate is simple the total minus the background rate.}
}
\end{flushleft}
\end{document}